\newcommand{\be}{\begin{equation}}\newcommand{\ee}{\end{equation}}
\newcommand{\bea}{\begin{eqnarray}}\newcommand{\eea}{\end{eqnarray}}
\newcommand{\nn}{\nonumber}
\newcommand{\w}{w_{1+\infty}}
\documentstyle[12pt]{article}

\begin{document}
\renewcommand{\thefootnote}{\fnsymbol{footnote}}
\thispagestyle{empty}
\begin{flushright} LNF-92/099 (P)\\
ENSLAPP-L-409/92 \\November 1992
\end{flushright}

\bigskip\begin{center}
{\large\bf MULTI-FIELD COSET SPACE REALIZATIONS OF $\w$}\vspace{0.6cm} \\
Stefano Bellucci\footnote{BITNET: BELLUCCI@IRMLNF} \vspace{0.3cm} \\
{\it INFN--Laboratori Nazionali di Frascati} \\
{\it P.O. Box 13, I-00044 Frascati, Italy} \vspace{0.3cm} \\
Evgenyi Ivanov\footnote{BITNET: EIVANOV@ENSLAPP.ENS-LYON.FR} \vspace{0.3cm} \\
{\it Laboratoire de Physique Theorique ENSLAPP,} \\
{\it ENS Lyon, 46 Allee d'Italie, F-69364 Lyon Cedex 07, France}
\vspace{0.3cm}\\
{\it and} \vspace{0.3cm} \\
{\it JINR-Laboratory of Theoretical Physics} \\
{\it Dubna, Head Post Office, P.O.Box 79, 101 000 Moscow, Russia}
 \vspace{0.3cm} \\
and \vspace{0.3cm} \\
Sergey Krivonos\footnote{BITNET: KRIVONOS@LTP.JINR.DUBNA.SU} \vspace{0.3cm}\\
{\it JINR-Laboratory of Theoretical Physics} \\
{\it Dubna, Head Post Office, P.O.Box 79, 101 000 Moscow, Russia}
\vspace{0.3cm}\\
{\bf Abstract}
\end{center}
We extend the coset space formulation of the one-field realization of
$\w$ to include more fields as the coset parameters. This can be done
either by choosing a smaller
stability subalgebra in the nonlinear realization of $\w$ symmetry, or by
considering a nonlinear realization of some extended symmetry, or by
combining both options. We show that all these possibilities
give rise to the multi-field realizations of $\w$. We deduce
the two-field realization of $\w$ proceeding from a coset
space of the symmetry group $\tilde{G}$ which is an extension of $\w$
by the second self-commuting set of higher spin currents. Next,
starting with the unextended $\w$ but placing
all its spin 2 generators into the coset, we obtain a new
two-field realization of $\w$ which essentially involves a $2D$ dilaton.
In order to construct the invariant action for this system we
add one more field and so get a new three-field realization of $\w$.
We re-derive it within the coset space approach, by applying the latter to
an extended symmetry group $\hat{G}$ which is a nonlinear deformation
of $\tilde{G}$. Finally we present some multi-field generalizations
of our three-field realization and
discuss several intriguing parallels with $N=2$ strings and
conformal affine Toda theories. \\

\vfill
\setcounter{page}0
\renewcommand{\thefootnote}{\arabic{footnote}}
\setcounter{footnote}0
\newpage

\section{Introduction}
A universal geometric description of field systems respecting the
invariance under a symmetry group $G$ is provided by the method
of nonlinear realizations (or the coset space realizations). One
considers $G$
as a group of transformations acting in some coset space $G/H$ with an
appropriately chosen stability subgroup $H$ and identifies the coset
parameters with fields. The authors of ref. \cite{a0}
have constructed a nonlinear realization of $\w$ and have shown that, after
imposing an infinite number of the inverse Higgs-type constraints on the
relevant Cartan forms, one is left with the
well-known realization of $\w$ on one scalar $2D$ field \cite{aa}.

In the present paper we extend the results of ref. \cite{a0} to obtain some
multi-field realizations of $\w$. The main idea of our approach is to
enlarge the initial coset, in order
to find the appropriate place for additional
scalar fields. This can be done in two ways, either by considering a
larger group $\tilde G$ which includes $\w$ as a subgroup\footnote{We denote
the algebra $\w$ and the associate group of transformations by the
same character, hoping that this will not give rise to a confusion.},
or by choosing a smaller stability subgroup $\tilde H$ in $\w$. We
demonstrate
that both options (and their combination) give rise to the multi-field
realizations of $\w$,
allowing us to obtain the  two-field realization of \cite{a2}, as well
as essentially new realizations
of $\w$ on the set of scalar fields including dilaton-like ones.
Besides providing new insights into the geometric origin of the
$\w$ transformations, this could shed more light on the geometry
of the associated $w$ gravity.

The paper is organized as follows.

In Sect.2 we briefly recall the main results of ref. \cite{a0} concerning
the one-field realization of $\w$.

In Sect.3 we recapitulate the basic facts about the two scalar field
realization of $\w$ and list some important symmetries of the
corresponding action. In particular, we
present an infinite number of conserved currents extending the
standard $\w$ transformations.

Our main results are collected in Sects.4 and 5.

In Sect.4 we utilize the symmetries presented in the previous section,
in order to recover the
two scalar field realization of $\w$ within the coset space approach.
It turns out that in this case one should start with
an extension of the algebra $\w$ by some infinite-dimensional ideal.
We also construct a new coset two-field realization of $\w$ which
essentially involves a $2D$ dilaton.

In Sect.5 we show that this $2D$ field system can be given a Lagrangian
formulation at cost of adding one more field. The resulting
three-field realization of $\w$ is in a sense interpolating between the
two-field ones constructed in Sect.4.
As in the previous cases, it can be deduced in the framework of the coset
space method and the
inverse Higgs procedure, now applied
to a nonlinear deformation of extended symmetry explored in Sect.4.
We explicitly give the relevant conserved currents and their OPE's
and discuss a possible relation to the
conformal affine Toda theory \cite{{a3},{a3a},{a3b}}. We also mention some
interesting
multi-field generalizations of the $\w$ realization constructed.

\section{A sketch of the one-field realization of $\w$}

For the reader's convenience we begin with recapitulating the basic steps
of the construction proposed in ref.\cite{a0}.

Its starting point
is the truncated $\w$ formed by
the generators $V^s_n$ with the following commutation relations
\be
\left[ V^s_n,V^k_m \right] = \left( (k+1)n-(s+1)m \right) V^{s+k}_{n+m};
\;\; s,k \geq -1;\;\; n \geq -s-1 ,m \geq -k-1 \; .\label{a1}
\ee
In what follows we will basically mean by $\w$ just this algebra.

As observed in ref. \cite{a0}, the standard one-field realization of
$\w$ \cite{aa} can be easily re-derived within the framework of a coset
realization.
It is induced by a left action of the group associated with the algebra
(\ref{a1}) on the infinite-dimensional coset over the
subgroup generated by
\be
V^0_n \; (n \geq 0) \quad , \quad V^s_m \;
(s \geq 1, m \geq -s-1) \; .\label{a2}
\ee
An element of this coset space  can be parametrized as
follows:
\be
g= e^{zV^0_{-1}}\; e^{v_0V^{-1}_0}\; e^{\sum_{n \geq 1} v_n V^{-1}_n}\; .
\label{a3}
\ee
Here all coset parameters are assumed to be $2D$ fields depending on $z$
and the second $2D$ Minkowski space light-cone coordinate which is regarded
as an extra parameter. In what follows we will never explicitly indicate the
dependence
on this extra coordinate and, where it is necessary, will identify
$z$ with the light-cone coordinate $x^{+}$.

As usual in nonlinear realizations, the group $G$ (associated with $\w$
in the present case) acts as
left multiplications of the coset element:
\be
g_0 \cdot g(z,v_0,v_n)=g(z',v_0',v_n') \cdot h (z,v_0,v_n;g_0) \quad , \quad
g_0=exp (\sum_{s,n} a^s_n V^s_n ) \;,\label{a4}
\ee
where $h$ is some induced transformation of the stability subgroup. This
generates a group motion on the coset: the coordinate $z$ together with
the infinite tower of coset fields $v_0(z),v_n (z)$ constitute a closed
set under the group action. For instance,
\begin{eqnarray}
\delta^s z & = & -a^s(z)(s+1) (v_1)^s \nn \\
\delta^s v_0 & = & -a^s(z) s (v_1)^{s+1}\quad \mbox{etc.} \quad,  \label{a5}
\end{eqnarray}
where
$$
a^s(z) = - \sum_{n \geq -s-1} a^s_n z^{n+s+1}\; .
$$
Thus one obtains the realization of $\w$ on the coordinate $z$ and  the
infinite number of coset fields $v_0,v_n$. At the
next step of this game
the inverse Higgs procedure \cite{a4}
becomes involved, in order to find the kinematic equations for
expressing the higher-order coset fields in terms of $v_0(z)$.
This can be
done by putting some covariant constraints on the Cartan forms.

The Cartan forms are introduced in the usual way
\be
g^{-1}dg = \sum_{s,n} \omega^s_n V^s_n \label{a6}
\ee
and are invariant by construction under the left action of $\w$ symmetry.
They can be easily evaluated using the commutation relations (\ref{a1}).
The first few ones are as follows:
\begin{eqnarray}
\omega^0_{-1} & = & dz \nn \\
\omega^{-1}_{0} & = & dv_0 -v_1 dz \label{a7} \\
\omega^{-1}_{1} & = & dv_1 -2 v_2 dz ,\quad \mbox{etc.} \nn
\end{eqnarray}
Note that the higher order forms, like $\omega^{-1}_{0}$ and $\omega^{-1}_{1}$,
contain the pieces linear in the relevant coset fields. Now, keeping
in mind the invariance of these forms, one may impose the manifestly
covariant inverse Higgs type constraints
\be
\omega^{-1}_{n} =0 \quad , \quad  (n \geq 0) \label{a8}
\ee
which can be looked upon as algebraic equations for expressing the
parameters-fields $v_n \; (n \geq 1)$ in terms of $v_0(z)$ and its
derivatives;
e.g., using eqs. (\ref{a7}) one finds the coset fields $v_1$ and $v_2$ to be
expressed by
\be
v_1 = \partial v_0 \quad , \quad v_2=\frac{1}{2} \partial v_1=
 \frac{1}{2}\partial^2 v_0 \quad .
\label{a9}
\ee
Finally, substituting the expression for $v_1$ in the transformations laws
for $z$ and $v_0$ (\ref{a5}) one gets
\begin{eqnarray}
\delta^s z & = & -a^s(z)(s+1) (\partial v_0)^s \nn \\
\delta^s v_0 & = & -a^s(z) s (\partial v_0)^{s+1}\;.  \label{a10}
\end{eqnarray}
The active form of these transformations is just the standard one-field
realization of $\w$:
\be
{\tilde\delta} v_0(z) \equiv v'_0(z)-v_0(z) = a^s(z) (\partial v_0 )^{s+1}
\quad .
\label{a11}
\ee

Thus, the realization of $\w$ on one scalar field can be deduced in
a purely geometric way in
the framework of the nonlinear realizations method. One of the most
intriguing questions which arise when trying to advance
this approach further is how to
incorporate in it  the multi-field realizations of $\w$.
It seems natural to extend the coset space by some new generators
(either from the stability subgroup or by
considering a larger group $\tilde G$).
We will consider these possibilities in Sects.4 and 5, after briefly
reviewing
in the next section the well-known two scalar fields realization of $\w$.

\setcounter{equation}0
\section{Two scalar fields realization of $\w$}

In this Section we briefly recall the
two scalar fields realization of the $\w$ algebra.

It is known that algebraically $W_{1+\infty}$ admits a contraction to
$\w$. So one may obtain the field realizations of $\w$ as a contraction of
those of
$W_{1+\infty}$. This has been done  in ref. \cite{a2}. The corresponding
realization of $\w$ looks as follows.

Let $v $ and $w$ be the scalar fields with the following OPE's:
\be
v (z_1) v (z_2) = 0 \quad , \quad
v (z_1) w (z_2) =
w (z_1) w (z_2) =  log(z_{12})  \; ; \quad z_{12} \equiv z_1-z_2 \;\;.
\label{b1}
\ee
Then, defining the
current $V^{(s)}$
\be
V^{(s)} (z) = (\partial v)^{s+1}\partial w -
\frac{1}{s+2}(\partial v )^{s+2} \label{b2}\;\;\;,
\ee
one can easily check that it satisfies the following OPE
\be \label{b2a}
V^{(s)}(z_1)V^{(k)}(z_2)= \frac{(s+k+2)V^{(s+k)}(z_2)}{z_{12}^2}
+\frac{(s+1)\partial V^{(s+k)}(z_2)}{z_{12}}-
\frac{ \delta_{s+1,0}\delta_{k+1,0} }{z_{12}^2}
\ee
from which it follows that the  Fourier components
\be
V^s_n = \frac{1}{2\pi i} \oint dz z^{s+n+1} V^{(s)} (z) \label{b3}
\ee
obey the commutation relations of $\w$, eq.(2.1)\footnote{The central
term appearing in the spin 1 sector of the OPE (\ref{b2a} )
does not contribute to the commutation relations
of the truncated $\w$ algebra (2.1) due to the restrictions on the
indices $n$, $m$.}.
The associate transformations of $v$ and $w$ read
\begin{eqnarray}
\delta^s v & = & a^s(z) (\partial v )^{s+1} \nn \\
\delta^s w & = & a^s(z) (s+1)(\partial v )^{s} \partial w \quad .\label{b4}
\end{eqnarray}
The free action for these fields
\be
S= \int d^2 z \left( -\partial_{+}v \partial_{-}v+
     \partial_{+}w \partial_{-}v  +
     \partial_{-}w \partial_{-}w \right) \label{b5}
\ee
gives the simplest example of a two-field action invariant
under $\w$ transformations. In fact the first term in the action is invariant
in its own right and can be removed by the redefinition of $w$
\begin{eqnarray}
w &\Rightarrow & \tilde{w}\;,\quad \tilde{w} = w - \frac{1}{2}v \label{b5a} \\
S &=& \int d^2 z \left( \partial_{+}\tilde{w}\partial_{-}v +
\partial_{-}\tilde{w}\partial_{+}v \right) \label{b5b} \\
v(z_1)\tilde{w}(z_2)&=& log (z_{12})\;,\quad
 \tilde{w}(z_1)\tilde{w}(z_2) = 0\;, \quad v(z_1)v(z_2) = 0\;. \label{b5c}
\end{eqnarray}

Let us make a few comments concerning the realization (\ref{b4}) and action
(\ref{b5}), (\ref{b5b}).

First of all we  note that the transformation law for the field $w$
in (\ref{b4}) can be rewritten as
\be
\delta^s w = -\left( \delta^s z \right) \partial w \;,\label{b6}
\ee
where $\delta^s z $ is given in (\ref{a10}). So the $\w$
transformations of the field
$w$ are induced by the $\w$ shift of its argument
$z$. Thus this field  behaves as a scalar under $\w$ symmetry, while
$v$ supports the standard one-field realization of
$\w$ discussed in the previous section.

Secondly, the free action (\ref{b5}), (\ref{b5b}) possesses
a larger symmetry than $\w$. Here we quote the infinite number of
conserved currents which generate the symmetries we will discuss in
the next Sections. These currents read as follows:
\begin{eqnarray}
W^{(s)}_1 & = & \frac{1}{s+2}\; (\partial v )^{s+2} \label{b7} \\
W^{(s)}_2 & = & (\partial v )^{s-1}\partial^3 v
\nn \\
W^{(s)}_3 & = & (\partial v )^{s-2}\partial^4 v \nn \\
& \ldots & \nn \\
W^{(s)}_N & = & (\partial v )^{s-N +1}\partial^{N+1}v \;,
\quad \mbox{etc.} \label{b7a}
\end{eqnarray}
For each $N$ there exists a value $s_0$ such that the currents with the
spins exceeding $s_0 + 2$ are independent in the sense that
they cannot be reduced to the derivatives of the lower $N$ currents.
Using the OPE's (\ref{b1}) one can check that all these currents
mutually commute and give rise to the transformations of $w$
only ($\delta v=0$):
\begin{eqnarray}
\delta^{s}_1 w & = & b^s(z) (\partial v)^{s+1} \nn \\
\delta^{s}_2 w & = & (s-1)b^s(z) (\partial v)^{s-2}
 \partial^3 v
+\partial^2\left( b^s(z)(\partial v)^{s-1} \right)
   \label{b8} \\
\delta^{s}_3 w & = &  (s-2)b^s(z) (\partial v)^{s-3}
 \partial^4 v
  -\partial^3 \left( b^s(z)(\partial v)^{s-2} \right) \; , \quad
\mbox{etc.} \nn
\end{eqnarray}

It is worth mentioning that the currents $W^{(s)}$ (\ref{b7}),
(\ref{b7a}) together
with the $\w$ currents $V^{(s)}$ (\ref{b2}) form a closed algebra.
Moreover, they span an ideal $\hat{\cal H}$ in this extended algebra,
so that
the factor-algebra of the latter by $\hat{\cal H}$ coincides with $\w$.
Let us give the OPE's between $V^{(s)}$ and the first two
currents $W^{(s)}_1,\; W^{(s)}_2$
\begin{eqnarray}
V^{ (s) }(z_1)W_1^{ (k) }(z_2) & = & \frac{(s+k+2)W_1^{(s+k)}(z_2)}{z_{12}^2}
     +\frac{(s+1)\partial W_1^{(s+k)}(z_2)} {z_{12}}
      +\frac{\delta_{s+1,0}\delta_{k+1,0} } {z_{12}^2} \nn \\
V^{(s)}(z_1)W_2^{(k)}(z_2) & = & (k-1)(s+1) \left[
\frac{W_1^{(s-1)}(z_1) \; W_2^{(k-1)}(z_2)}{z_{12}^2}
     +6\frac{W_1^{(s-1)}(z_1) \; W_1^{(k-3)}(z_2)}{z_{12}^4} \right]\nn \\
       & = & A_1(s,k) \frac{ W_1^{(s+k-2)}(z_2) } {z_{12}^4} + \ldots
     + A_2(s,k) \frac{W_2^{(s+k)}(z_2)}{z_{12}^2} + \ldots  \nn \\
      & & + A_3(s,k) \frac{W_3^{(s+k+1)}(z_2)}{z_{12}}\;, \label{b8a}
\end{eqnarray}
where in the last two lines we have written down only the leading terms
without specifying the numerical coefficients $A_1$, $A_2$, $A_3$
(actually in what follows we will never need their explicit form).
The OPE's between $V^{(s)}$ and the next currents display a similar structure.
Their  most
characteristic feature is that the currents $W^{(s)}_N$ form a not completely
reducible
set with respect to  $\w$: $W^{(s)}_1$ are transformed through themselves
while
the remaining  currents are transformed through themselves and the currents
$W^{(s)}_1$. So the minimal extended set of currents forming a closed
algebra includes $V^{(s)}$ and $W^{(s)}_1$. In what follows this minimal
extension
of $\w$ will be referred to as $\tilde{\cal G}$. Actually one can show
that adding any other current $W^{(s)}_N$, $N\geq 2$, to $\tilde{\cal G}$
would produce, via commuting with the $\w$ generators, the whole ideal
$\hat{{\cal H}}$. In the next Section
we will show that the two-field realization of $\w$ which we are discussing
here can be deduced from a coset realization of the group associated
with $\tilde{\cal G}$. Note that the ideal $\hat{{\cal H}}$
is none other than the universal enveloping algebra for the centreless
$U(1)$ Kac-Moody algebra generated by the spin 1 current
$W_1^{(-1)}$. Indeed, this enveloping algebra is spanned by all possible
products of $\partial v$ and its derivatives of any order.
All such products can be represented as linear combinations
of the currents $W_N^{(s)}$ from the set (\ref{b7}), (\ref{b7a}) and the
derivatives
of $W_N^{(s)}$, so these currents form a basis in the enveloping algebra
in question regarded as a linear algebra.

The existence of extra symmetries in the above system entails an
interesting consequence. One
can modify the currents $V^{(s)}$ by adding $W_1^{(s)}$ with the proper
coefficient so as the modified currents still close on $\w$. This
modification reads as follows:
\begin{equation}
V^{(s)}_{(\gamma)} = V^{(s)} + \gamma s W^{(s)}_1\;, \label{b8b}
\end{equation}
with $\gamma$ being an arbitrary parameter. Thus, in the case at hand
one actually deals with a one-parameter family of $\w$ algebras.
This fact has been noticed in ref. \cite{a5} where it has been also
observed  that, by adjusting the parameter value, one can cancel the
central charge in the spin 1 sector of $\w$ and so make the
modified currents obey the centreless $\w$.
As it is seen from the OPE's (\ref{b2a}) and (\ref{b8a}), in our notation this
cancellation occurs at
$\gamma = -\frac{1}{2}$. It is easy to show that
$$V^{(-1)}_{(-1/2)} = \partial \tilde{w}\;,$$
and the spin 1 current becomes self-commuting\footnote{The
central term is still retained
in the commutator (or OPE) of $V^{(-1)}_{(-1/2)}$ with $W^{(-1)}_1$.}
as a consequence of the self-commutativity
of the field $\tilde{w}$.

For any value of $\gamma$ one can pick up the combination of the fields
$w$, $v$, namely $w + \gamma v$, which transforms under the appropriate
$\w$ from the above family according to the transformation law
(\ref{b6}) (the transformations
of $v$ do not depend on $\gamma$ and are always given by eq.(\ref{b4})).
In particular, for the centreless $\w$ (for $\gamma=-\frac{1}{2}$) this
combination coincides with the self-commuting field $\tilde{w}$
\begin{equation}
V^{(s)}_{(-1/2)}: \quad \delta \tilde{w} = -(\delta^s z)\partial \tilde{w}\;.
\end{equation}

Finally, we note that the action (3.6), (3.8) actually respects many more
symmetries than those listed above. In particular, in view of invariance
of (3.8) under the permutation $\tilde{w} \Leftrightarrow v$, the currents
obtained by this permutation from (3.2), (3.11), (3.12), (3.15) also
define symmetries of (3.6), (3.8). We discussed only those symmetries
which are relevant to the subsequent coset space constructions.

\setcounter{equation}0
\section{Two-field realizations of $\w$ from the coset space approach}

In this Section we explain how the two-field realizations
of $\w$ reviewed in the previous Section can be reproduced in the framework of
the coset space approach and also present a new kind of
two-field realization.

As we noted previously, in order to find the appropriate place for the
additional fields we need to enlarge the coset
space we start with.
The first possibility is to enlarge the symmetry group $G$ (as the
simplest  example we may consider the nonlinear realization of the extended
symmetry $\tilde G$ defined in the previous Section). We do this in Subsect.
4.1
and demonstrate that the two-field realizations of ref. \cite{{a2},{a5}}
can be derived in this way. Secondly, one may still deal with the same
$\w$ as in ref. \cite{a0}, but restrict the stability subgroup,
transferring some of its generators in the coset. We do this in Subsect. 4.2
and deduce a new example of the two-field realization of $\w$.

\subsection{Nonlinear realization of the extended symmetry $\tilde G$}

Let us consider the algebra $\tilde{{\cal  G}}$ with the following commutation
relations:\footnote{This algebra can be regarded as a contraction of the
sum of two independent $\w$ algebras.}
\begin{eqnarray}
\left[ V_{n}^s,V_{m}^k \right] & = & \left( n(k+1) - m(s+1) \right)
                 V_{n+m}^{s+k} \nn \\
\left[ W_{n}^s,V_{m}^k \right] & = & \left( n(k+1) - m(s+1) \right)
                 W_{n+m}^{s+k} \label{c1} \\
\left[ W_{n}^s,W_{m}^k \right] & = & 0 \nn
\end{eqnarray}
The generators $V^s_n$ give rise to the standard $\w$ algebra and the
mutually commuting generators $W^s_n$ are in the adjoint representation
of $\w$. They are none other
than the Fourier modes of the currents $V^{(s)}$ and $W^{(s)}_1$ introduced in
the previous Section (eqs. (3.2) and (\ref{b7})).
 From eqs.(\ref{c1}) it is evident
that $W_{n}^s$ constitute an infinite-dimensional ideal in $\tilde{\cal G}$.

As discussed in Section 2, for constructing a nonlinear realization
of the associated symmetry group $\tilde G$ one needs to define
the appropriate infinite-dimensional coset space with a suitably chosen
stability subgroup $H$. In the present case there are many
possible choices for the stability subgroup, due to the commutativity
of the generators $W_n^s$. Here we consider the simplest possibility, with
the stability subalgebra formed by the following generators:
\be
V^0_n \; (n \geq 0) \quad , \quad V^s_m \; (s \geq 1) \; ; \;
W^k_n \; (k \geq 0) \; .\label{c2}
\ee
An element of the associate coset space  can be parametrized as
follows:
\be
g= e^{zV^0_{-1}}\;  e^{\sum_{n \geq 0} v_n V^{-1}_n}\;
 e^{\sum_{m \geq 0} w_m W^{-1}_m}\; .
\label{c3}
\ee
As usual, the group $\tilde G$ acts as the left multiplications
on the coset element (we will first consider the action of
the $\w$ subgroup of $\tilde G$ with the generators $V^s_n$):
\be
g_0 \cdot g(z,v_n,w_m)=g(z',v_n',w'_m)\cdot h (z,v_n,w_m;g_0) \quad , \quad
g_0=exp (\sum_{n\geq -s-1} a^s_n V^s_n ) \quad .\label{c4}
\ee
Now the coordinate $z$ constitutes a closed set under the group action
together with
the infinite tower of coset fields $v_n (z),w_m (z)$:
\begin{eqnarray}
\delta^s z & = & -a^s(z)(s+1) (v_1)^s \nn \\
\delta^s v_0 & = & -a^s(z) s (v_1)^{s+1}  \label{c5} \\
\delta^s w_0 & = & 0\;,   \quad \mbox{etc.}\;,        \nn
\end{eqnarray}
where
$$
a^s(z) = - \sum_{n \geq -s-1} a^s_n z^{n+s+1}\; .
$$
Thus, we obtained the realization of $\w$ on the coordinate $z$ and an
infinite number of coset fields $v_n,w_m$.
Now we may impose the inverse Higgs constraints
in order to find the kinematic equations for
expressing the higher-order coset fields in terms of $v_0(z),w_0(z)$.
The appropriate set of constraints reads as follows:
\be
\omega^{-1}_{n} =0 \quad , \quad
{\tilde\omega}^{-1}_{n} =0 \quad , \quad  (n \geq 0) \label{c6}
\ee
where $\omega^s_n,{\tilde\omega}^k_m$ are Cartan forms:
\be
g^{-1}dg = \sum_{s,n} \omega^s_n V^s_n
+ \sum_{k,m} {\tilde\omega}^k_m W^k_m \quad .
\label{c7}
\ee
After straightforward calculations one finds that the first higher-order coset
fields
$v_1,\;w_1$ are expressed by
\be
v_1 = \partial v_0 \quad , \quad w_1= \partial w_0 \quad .
\label{c8}
\ee
Finally, substituting the expression for $v_1$ in the transformations laws
for $z$ and $v_0$ (\ref{c5}), we obtain
\begin{eqnarray}
\delta^s z & = & -a^s(z)(s+1) (\partial v_0)^s \nn \\
\delta^s v_0 & = & -a^s(z) s (\partial v_0)^{s+1}  \label{c9} \\
\delta^s w_0 & = & 0 \quad . \nn
\end{eqnarray}
These transformations, being rewritten in the active form, are
recognized as the standard two-field realization of $\w$ (cf. eqs.(\ref{b4}))
\begin{eqnarray}
{\tilde\delta} v_0(z) & \equiv & v'_0(z)-v_0(z)
  = a^s(z) (\partial v_0 )^{s+1} \nn \\
{\tilde\delta} w_0(z) & \equiv & w'_0(z)-w_0(z)
  = a^s(z) (s+1) (\partial v_0 )^{s}\partial w_0 \quad .\label{c10}
\end{eqnarray}

As for the transformations with the generators
$W^s_n $,
they do not touch the coordinate and the parameter-fields associated
with the generators $V^{-1}_n$ and act only on $w_m (z)$. In particular,
\be
\delta^s w_0 = b^s(z) (v_1)^{s+1} = b^s (z)(\partial v_0)^{s+1}\;.\label{c11}
\ee

In accordance with the remark in the end of Sect.3, there exists a
one-parameter
family of embeddings of $\w$ in $\tilde{G}$. Introducing the new basis in
$\tilde{G}$
\be
V^{s}_{(\gamma)n} = V^{s}_n +\gamma s W^{s}_n\;, \quad W^s_n\;, \label{c12}
\ee
one easily checks that the newly defined generators satisfy the same
commutation relations as the old ones. In order to study the action of the
$\w$ transformations with a given $\gamma$ on the coset fields, it is
convenient to pass in eq. (\ref{c3}) to the new set of generators
(\ref{c12}). This entails an appropriate  redefinition of the coset
fields, on the new set of fields the $\w^{(\gamma)}$ transformations
being realized just as the original $\w$ transformations (corresponding to
$\gamma = 0$) on $v_m$ and $w_n$. It is easy to check that the field
$v_0$ is not redefined while the generator $W^{-1}_0$ now enters with the
following combination of original fields:
$$ w_0 + \gamma v_0 .$$
The relevance of such a combination has been already mentioned in the end
of Sect.3. Here we see how it appears within the coset space approach.

Thus, we have succeeded in deducing the standard realizations of $\w$
on two scalar fields in the framework of the nonlinear realizations method.
One of these fields, $v$, as before is the coset field associated
with the generator $V^{-1}_0$ from the spin 1 sector of $\w$, while
the other, $w$ (or $\tilde{w}$), is associated with the
generator $W^{-1}_0$ from the spin 1 sector of the infinite-dimensional
ideal $\tilde{{\cal H}}$ extending $\w$ to $\tilde{{\cal G}}$. It is
worth mentioning that the same fields can be alternatively interpreted as the
coset parameters corresponding to a nonlinear realization of the symmetry
isomorphic to $\tilde{{\cal G}}$ and generated by the currents
following from (3.2), (3.11) via the change
$\tilde{w} \Leftrightarrow v$. With respect to this symmetry, the roles of
the fields $\tilde{w}, \; v$ are inversed: $\tilde{w}$
comes out as the coset parameter related to the subalgebra $\w$ while
$v$ is associated with the ideal.

\subsection{Nonlinear realization of $\w$ with two essential fields}

Now we consider another possibility to obtain a two-field realization
of $\w$. This realization turns out to be of an
essentially new kind compared to the previously known ones.

The starting point of our construction will be the
realization of $\w$
in the coset space $\w /H$, where the stability subgroup $H$ is now
generated by
\be
 V^s_m \; (s \geq 1) \; .\label{c13}
\ee
In other words, we put in the coset space along with the spin 1 generators
all the generators from the (truncated) Virasoro subalgebra in the spirit
of ref.
\cite{{ab},{a7},{ac}}.

An element of this coset space can be parametrized as
follows:
\be
g= e^{zV^0_{-1}}\;  e^{\sum_{n \geq 0} v_n V^{-1}_n}\;
   e^{\sum_{m \geq 0} u_m V^{0}_m}\; .
\label{c14}
\ee
Thus we have now two infinite series of the coset parameters-fields
associated with
the generators $V^{-1}_n$ and $V^{0}_m$.

As in the previous cases we may easily find the transformation properties of
the fields $v_n,\;u_m$ under the $\w$ symmetry realized by left multiplications
of the coset (\ref{c14})

\begin{eqnarray}
\delta^s z & = & -(s+1)\; a^s \; (v_1)^s \nn \\
\delta^s v_0 & = & -s \; a^s \; (v_1)^{s+1}  \label{c15} \\
\delta^s u_0 & = & -(s+1)\; \partial a^s \; (v_1)^s -
  2s \; a^s \; (v_1)^{s-1} v_2\;\;,  \quad \mbox{etc.}        \nn
\end{eqnarray}
Using the inverse Higgs-type constraints on the Cartan forms
$$g^{-1}dg = \sum \omega^s_n V^s_n$$
\be
\omega^{-1}_{n} =  \omega^{0}_{n} = 0 \quad , \quad  (n \geq 0)
\label{c16}
\ee
one finds that all the coset fields are covariantly expressed through the two
independent ones, $v_0$ and $u_0$. In particular, the coset fields
$v_1$ and $v_2$ are represented by
\be
v_1 = \partial v_0 \quad , \quad v_2=\frac{1}{2} \partial v_1=
   \frac{1}{2}\partial^2 v_0 \quad .
\label{c17}
\ee
Finally, substituting the expressions for $v_1,v_2$ in the transformations laws
of $z,v$ and $u$ (\ref{c15}), we find the  new
two-field realization of $\w$
\begin{eqnarray}
\delta^s z & = & -(s+1)\; a^s \; (\partial v_0)^s \nn \\
\delta^s v_0 & = & -s \; a^s \; (\partial v_0)^{s+1}  \label{c18}\\
\delta^s u_0 & = & -(s+1)\;\partial \left( a^s \; (\partial v_0)^{s}\right) \nn
\end{eqnarray}
or, in the active form
\begin{eqnarray}
{\tilde\delta^s} v_0 & = & a^s \; (\partial v_0 )^{s+1} \nn \\
{\tilde\delta^s} u_0 & =& -(s+1) \; \partial \left( a^s \;(\partial
v_0)^{s}\right)
 +(s+1)\; a^s \; (\partial v_0 )^{s}\partial u_0 \quad .\label{c19}
\end{eqnarray}

In this new realization the field $v_0$ and the coordinate $z$
form as before a representation of $\w$ in their own, while  $u_0$ transforms
like the dilaton in the standard conformal theories. To see this, it
is instructive to rewrite the transformation law of $u_0$ in the form
\be \label{c19a}
{\tilde\delta^s} u_0 = \partial (\delta^s z) - \delta^s z \;\partial u_0\;.
\ee
Under the conformal group ($s=0$) $u_0$ transforms as a $2D$ dilaton, with
the purely chiral parameter $\delta^0 z = -a^0 (z)$. However, for the higher
spin transformations $\delta^s z$ ceases to be chiral because of the
explicit presence of $\partial v_0$, so the $\w$ transformations
of $u_0$ can be reduced to the conformal ones only provided the field $v_0$ is
on shell.

Several comments are in order here.

First, it should be emphasized that the second realization can be
straightforwardly extended by adding more generators (with spins 3, 4, etc.)
to the coset. In
this case we will obtain some  multi-field realizations of $\w$. Moreover,
we are at freedom to combine the approaches used in the
Subsections 4.1 and 4.2, including more generators
in the coset as well as adding more currents, in order to get new
realizations of $\w$. This possibility will be exploited in the next Section.

As a second remark closely related to the first one, we note that including
in this game the generators with higher spins $(s \geq 3)$ immediately
leads to the appearance of the pure shifts of the corresponding fields under
higher spin transformations (due to the goldstone nature of these fields).
Classically,
if these systems admit any Lagrangian formulation, such shifts may appear
only through Feigin-Fuchs-like terms in the relevant
currents. At the same time,  these terms would inevitably yield central charges
in the commutators of the higher spin currents. On the other hand,
the algebra $\w$ is known to admit central extensions in the spin 1
and spin 2 sectors only, otherwise it should be deformed into $W_{1+\infty}$.
So the
nonvanishing central charges in the commutators of the higher spin currents
could signal an existence of a hidden  $W_{1+\infty}$ structure in these
systems. Another, more prosaic solution to this controversy could consist in
that these systems admit no Lagrangian formulations at all (see the next
comment) and the only correct way to include the higher spin generators
in the cosets is to start from the beginning from nonlinear realizations
of $W_{1+\infty}$ or to extend $\w$ in a proper way. We postpone the
complete analysis of this interesting question
to the future.

Finally, we wish to emphasize that any new field realization
of $\w$ (or of $W_{1+\infty}$) seems to make sense only if it follows
from some Lagrangian system (at least from the system of free fields).
But just in the second, dilaton-like realization  we did not succeed in
finding any meaningful Lagrangian possessing
invariance under the transformations (\ref{c19}). Respectively,
it seems impossible to define currents which would generate (\ref{c19})
via Poisson brackets\footnote{The sum of the free actions for the fields
$u_0$ and $v_0$ is invariant under the following modified
transformations: $u_0$ is still transformed according to
(\ref{c19}) while $v_0$ according to
$$
\tilde \delta v_0 = a^s(z) (\partial v_0)^{s+1} + s(s+1) a^s(z)
(\partial v_0)^{s-1} (\partial^2 u_0 +
\frac{1}{2} (\partial u_0)^2\;)\;.
$$
However, the Lie bracket structure of these transformations is not that of
$\w$. Rather this is a kind of $W_{1+\infty}$ symmetry analogous to the
one found by two of us (S.B.\& E.I.) in
the system of two fields, one of which enters with a Liouville potential term
\cite{a6}. Note that the Liouville term for $u_0$, $exp{(-u_0)}$, is in itself
invariant under the above transformations (modulo a total derivative).}.
However, in a funny way it turns out that such a
Lagrangian can be constructed if one adds to the system $\{ v_0,\; u_0  \}$
one more field, with a specific transformation law under $\w$. We
discuss this possibility in the next Section.

\setcounter{equation}{0}
\section{New three-field coset space realization of $\w$}

In this Section we consider a  more complicated coset realization of $\w$.
It is based on a combination of the two approaches used in the previous
Section.
The main idea is to start with some extended group
$\hat{G}$ which contains $\w$ as a factor group. We first present the
invariant Lagrangian and the transformation laws for the relevant field system
and then explain how these laws can be derived from the coset
space formalism.

Let us consider the following three-field modification of the action (3.6),
(3.8)
\begin{eqnarray}
S &=& \int d^2 z \left( -\partial_{+} v \partial_{-} v +
\partial_{+} w \partial_{-} v +\partial_{-} w \partial_{+} v +
 \partial_{+} u \partial_{-} u
 \right) \nn \\
& = & \int d^2 z \left( \partial_{+} \tilde{w} \partial_{-}v +
\partial_{-} \tilde{w} \partial_{+}v + \partial_{+}u \partial_{-}u
\right)  \label{d6} \quad .
\end{eqnarray}
The latter is invariant under the transformations
\begin{eqnarray}
{\tilde\delta}^s v & = & a^s(z) (\partial v)^{s+1} \nn \\
{\tilde\delta}^s u & =& -\alpha
\partial \left( a^s(z) (s+1)(\partial v)^{s}\right)
 +a^s(z)(s+1) (\partial v)^{s}\partial u \label{d3} \\
{\tilde\delta}^s w & =& \frac{1}{2}s(s+1) a^s(z)(\partial v)^{s-1}
\left( (\partial u)^2 +2\alpha\partial^2 u \right) \nn \\
 & + & a^s(z)(s+1) (\partial v)^{s}\partial w  \nn \quad ,
\end{eqnarray}
with $\alpha$ being an arbitrary parameter. One sees that
the transformations
of $v$ and $u$ coincide with (\ref{c19}) (the parameter $\alpha$
can be introduced in (\ref{c19}) via a rescaling of $u$) and so constitute
$\w$ with the classical central charge $\sim \alpha^2$ in the Virasoro
sector. Thus, for these fields we still have the dilaton-like
realization of $\w$ constructed in Subsect. 4.2 (for the moment we
discard the subscript $0$ of these fields). However, the above
action is invariant only due to the presence of the third
field $w$ (or $\tilde{w}$). Its transformation law for $s\geq 1$
essentially differs from
(3.5) and can by no means be reduced to the latter, since putting $u$ or
the stress-tensor of $u$, $T(u) = (\partial u)^2 + 2\alpha \partial^2 u $,
equal to
zero would clearly break the $\w$ symmetry.

To understand why this
reduction does not work, one needs to consider a commutator of two $\w$
transformations on $w$. One finds that the commutator of two such
transformations,
with the spins $s+2$ and $k+2$, besides the
original $\w$ transformation of $w$ with the spin $s+k+2$ and the bracket
parameter
 $$
a^{s+k}_{br} = (s+1) a_1^s\partial a^k_2 - (1 \leftrightarrow
2,\; s \leftrightarrow k)\;,
$$
necessarily contains some extra terms
proportional to $\alpha^2$. They are none other than the transformations from
the set (3.13), not only those corresponding to the variation
$\delta^s_1$, but also those associated with higher order variations. Their
contribution appears already in the commutator of the spin 2 and spin 3
$\w$ variations (a pure chiral shift of $w$ generated by the spin 1 current
$W^{(-1)}_1$ ), so the spin 3 current in the present case turns out to be
not primary. Commuting these extra variations of $w$ with the
original ones (\ref{d3}) we actually produce all the transformations
from the set (3.13) which, as mentioned in Sect. 3 (see
discussion below eq. (3.14)), form an infinite-dimensional ideal
$\hat{\cal H}$ coinciding with the enveloping algebra of the spin 1 current
$W_1^{(-1)}$. Thus,
on the field $w$ the $\w$ symmetry is defined only modulo the
ideal ${\hat {\cal H}}$ formed by the mutually commuting transformations
(3.13). In other words,
$\w$ now appears as a factor algebra of an extended algebra
${\hat {\cal G}}$ by the ideal ${\hat {\cal H}}$. The transformations from
this ideal are realized only on the field $w$, which explains why
on the fields $v$ and $u$ the extended symmetry is reduced to $\w$. For the
time being,
we do not completely understand
why we need the additional field $w$ and the factor-algebra interpretation
of $\w$ in order to construct the invariant action for the system $v,\;u$.

For a better insight into the structure of $\hat{{\cal G}}$ and in order
to learn how to regain the above three-field realization from the coset
space approach, we should construct the relevant currents and compute
their OPE's, proceeding from the canonical formalism corresponding to the
action (\ref{d6}).

The $\w$ currents are easily found to be
\begin{eqnarray}
{\hat V}^{(s)} & = & (\partial v)^{s+1}\partial w -\frac{1}{s+2}
(\partial v)^{s+2}
+(s+1)(\partial v)^s \left(\frac{1}{2} (\partial u)^2
+\alpha\partial^2 u \right) \label{d5} \quad .
\end{eqnarray}
They generate the transformations (\ref{d3}) via the canonical OPE's
\be
w(z_1)v(z_2)=w(z_1)w(z_2)=u(z_1)u(z_2)=log (z_{12})\;,\quad v(z_1)v(z_2) = 0
\label{d4}
\ee
following from the action (\ref{d6}) (or, equivalently, via the canonical
Poisson brackets).
The currents ${\hat W}^{(s)}_1$ and the higher-order currents generating
the $\hat{{\cal H}}$ transformations (3.13) are still given by
the expressions (\ref{b7}) and (\ref{b7a}).

Now it is easy to write the whole set of OPE's defining the algebra
$\hat{{\cal G}}$
{}\footnote{We prefer to write here the OPE's instead of the
commutation relations, as the latter look rather
intricated for ${\hat {\cal G}}$.}
\begin{eqnarray}
{\hat V}^{(s)}(z_1){\hat V}^{(k)}(z_2) & = &
\frac{ (s+k+2) {\hat V}^{(s+k)} (z_2) }{ z_{12}^2 } +
\frac{ (s+1)\partial {\hat V}^{(s+k)}(z_2) }{ z_{12} }
-6 \alpha^2 \frac{ \delta_{s,0} \delta_{k,0} }{ z_{12}^4 } \nn \\
& - &
 6 \alpha^2 (s^2+s)(k^2+k)
\frac{ {\hat W}^{(s-2)}_1(z_1) {\hat W}^{(k-2)}_1(z_2) }{ z_{12}^4 }
-\frac{ \delta_{s+1,0} \delta_{k+1,0} }{z_{12}^2} \nn \\
{\hat V}^{(s)}(z_1){\hat W}^{(k)}_1(z_2) & = &
\frac{ (s+k+2) {\hat W}^{(s+k)}_1 (z_2) }{ z_{12}^2 } +
\frac{ (s+1)\partial {\hat W}^{(s+k)}_1(z_2) }{ z_{12} }+
\frac{ \delta_{s+1,0}\delta_{k+1,0} }{ z_{12}^2 }\nn \\
{\hat W}^{(s)}_1(z_1){\hat W}^{(k)}_1(z_2) & = & 0  \label{d1} \quad .
\end{eqnarray}

The main difference between ${\hat {\cal G}}$ and $\tilde{{\cal G}}$
considered in Sect.4
(recall that $\tilde{{\cal G}}$ is defined by the OPE (3.3) and by the first
OPE in (3.14)) is the appearance of
the bilinears of the currents
${\hat W}^{(s)}_1$ (along with the currents ${\hat W}^{(s)}_1$ themselves)
in the OPE of
${\hat V}^{(s)}(z_1){\hat V}^{(s)}(z_2)$. Hence $\hat{{\cal G}}$ is
a kind of {\it nonlinear} deformation of $\tilde{{\cal G}}$. From the above
OPE's it is clear
(after re-expanding the bilinears in the r.h.s. of eq. (5.5) over the
local products given at the point $z_2$ ) that actually not only the
currents $\hat{W}^{(s)}_1$
and their
products, but also the products involving derivatives of $\hat{W}_1^{(s)}$
appear.
Recalling that all such products can be represented as linear combinations
of the currents (3.11), (3.12) and the derivatives of the latter and that
these combinations constitute the ideal $\hat{{\cal H}}$ which coincides
with the enveloping algebra of $\hat{W}^{(-1)}_1$, one concludes that
${\hat {\cal G}}$ admits a twofold interpretation. Namely, it can be viewed
either as a nonlinear deformation of $\tilde{\cal G}$ with the same two
sets of
defining currents $\hat{V}^{(s)}$, $\hat{W}^{(s)}_1$ or as a linear
algebra generated by $\hat{V}^{(s)}$ and the infinite sequence of
currents $\hat{W}^{(s)}_1$, $\hat{W}^{(s)}_2$, ... $\hat{W}^{(s)}_N$, ... {}.

It is easy to figure out from the above OPE's
that the currents ${\hat W}$ indeed form
an ideal ${\hat {\cal H}}$ in ${\hat {\cal G}}$,
so that the factor-algebra of ${\hat {\cal G}}$ by ${\hat {\cal H}}$ is $\w$.
As before,
there exists a one-parameter family of embeddings of this $\w$ into
${\hat {\cal G}}$: one can check that the redefinition of ${\hat V}^{(s)}$
as in eq. (\ref{b8b}) does not affect the OPE's (\ref{d1}) except for the
central
terms where there appears a dependence on the parameter $\gamma$, such
that the central charge in the spin 1 sector of $\w$ vanishes at $\gamma =
-\frac{1}{2}$.   Note that
the closure of the Jacobi identities for ${\hat {\cal G}}$ could be
straightforwardly
checked  starting with its defining OPE's. But there is no actual need to do
this, since we have derived these OPE's  by specializing
to the field model (\ref{d6}) and making use of
the canonical formalism for the involved fields.

The interpretation of ${\hat {\cal G}}$ as a nonlinear algebra makes a bit
tricky the construction of the associated nonlinear realization.
Nonetheless, this can be
done using a proposal
of ref. \cite{a7} which seems to work for any nonlinear algebras. Namely,
one treats all the bilinears of the basic currents appearing in
the defining OPE's (or the bilinears of the defining
generators, if the standard commutation relations are used) as some new
independent objects, thus formally replacing the original nonlinear
algebra by some huge linear one which can already be handled by the
standard methods of coset realizations. Normally one puts all these new
currents (or generators) into the stability subalgebra so that
only the original, basic generators turn out to be actually
involved in the relevant coset constructions.

We will make use of all these ideas to construct a nonlinear realization
of the symmetry associated with ${\hat {\cal G}}$. In the case at
hand, the trick proposed in \cite{a7} corresponds just to sticking to the
interpretation of ${\hat {\cal G}}$ as a linear algebra involving as a
subalgebra the
whole enveloping algebra of the spin 1 current $\hat{W}_1^{(-1)}$. We
combine the two approaches used in Sect.3,
namely we put in the coset the generators with spin 1 and spin 2
from the set of ${\hat V}^{(s)}$, as well as the spin 1 generators
coming from ${\hat W}^{(k)}_1$. All the generators coming from the composite
currents (or, equivalently, from $\hat{W}^{(s)}_N$ for $N\geq 2$) are
placed in the stability subalgebra.

Thus, the representative of our coset reads
\be
{\hat g}= e^{z{\hat V}^0_{-1}}\; e^{\sum_{n \geq 0} v_n {\hat V}^{-1}_n}\;
  e^{\sum_{m \geq 0} w_m W^{-1}_m}\;
  e^{\sum_{m \geq 0} u_m {\hat V}^{0}_m}\; .
\label{d2}
\ee
The machinery for invoking the inverse Higgs phenomenon is the same
as in the previous Sections. After not very difficult labour  we find
that the whole symmetry
$\hat{{\cal G}}$ can be realized on
the three  essential scalar fields $v_0,u_0$
and $w_0$, whose transformations under the action of the $\w$ generators
are just those given by the equations (\ref{d3}). Thus we have succeeded in
deriving
the above three-field realization of $\w$ in a purely geometric way, proceeding
from the coset space of the associated extended symmetry $\hat{G}$.

Before ending this Section, let us make a few remarks concerning possible
generalizations of the above three-field realization and its relation
to other models.

First of all, let us note that in our realization the field $u$ enters in the
currents
and in the transformations of the field $w$  only through
its stress-tensor $T(u)$. Thus, we may
easily extend our realization, in a close analogy with the
$W_3$ realizations \cite{a8}, to the
multi-scalar case, replacing $T(u)$ by the corresponding stress-tensor
$$
T(u) \rightarrow T(u,\phi ,\ldots ) = T(u) + (\partial \phi)^2
+ \ldots  \quad ,
$$
where the fields $\phi, \ldots$ transform like $u$, but without
inhomogeneous pieces. For the moment it is not quite clear how to
reproduce such multi-field realizations in the framework of the
coset space approach. One could interpret the extra fields
$\phi,\;\ldots$ as scalars of ${\hat G}$ because these actually transform
only due to the $\w$ shift of one of their arguments (recall eq.
(\ref{c19a})), but then it is unclear how to ensure the appearance of
these fields in the transformation law of $w$ (where they enter through the
modified
stress-tensor $T(u, \phi , \ldots)$) that is necessary for the invariance of
the action. These reasonings prompt that the fields $\phi, \ldots$ should
be somehow incorporated from the beginning in the coset space approach
as the coset parameters.

The most intriguing possibility in what concerns these multi-scalar
realizations seems to be as follows. Let us consider the realization with one
extra
field $\phi$. One may add its kinetic term  to the action (\ref{d6}) with
the sign opposite to the sign of the kinetic term of $u$, assume that $\phi$
transforms
under $\w$ by an inhomogeneous law similar to the transformation
law of $u$ and,
respectively, add the appropriate Feigin-Fuchs term for $\phi$ to the
stress-tensor. Then, due to the wrong sign of the kinetic term for
$\phi$, the contributions of $u$ and $\phi$ to the central charge of
the classical stress-tensor can be cancelled (one may equivalently add
$\phi$ to the action with the normal kinetic term, but add a purely
imaginary Feigin-Fuchs term for this field to the stress-tensor).
Recalling that the $\hat{W}^{(s)}_N$ transformations which appear
in the commutators of the $\w$ variations of $w$ (and, respectively,
the $\hat{W}$ terms in the first of the OPE's (5.5)) are proportional to
the Virasoro central charge, the resulting transformations close on $\w$
without any
contributions from the $\hat{W}^{(s)}$ transformations and we end up
with the original $\w$ symmetry, now realized on four fields with the
signature $(2+2)$ in the target space. More precisely, this realization
is obtained by changing
$$
T(u) \rightarrow T(u, \phi) = T(u) - (\partial \phi)^2 + 2\alpha \partial^2
\phi
$$
in the transformation laws (5.2) and ascribing the following transformation
law to $\phi$
$$
\delta^s \phi = \alpha \partial ( a^s(z) (s+1)(\partial v)^s) +
a^s(z)(s+1)(\partial v)^{s}\partial \phi \;.
$$
This $(2+2)$ system deserves a further
study in view of its possible relation to the $N=2$ string of
Ooguri and Vafa \cite{a9}. A closely related observation
is that it naturally appears as the bosonic subsystem in the
following $N=2$ supersymmetric extension of
the $(1+1)$ action (\ref{b5}), (\ref{b5b}): one can $N=2$
supersymmetrize this action by introducing two $N=2$ chiral superfields
with the kinetic terms of the opposite sign. Then one may hope that
the $(2+2)$ system in question could be given a coset space interpretation
in the
framework of the nonlinear realizations of $N=2$ super
$\w$ algebras \cite{{a10},{a5}}, along the lines of \cite{a0} and the
present paper.

Finally, we wish to point out that our three-field system action (5.1)
coincides with the free part of
the action for the conformal affine Toda (CAT) system \cite{a3}, e.g.
under the identification
$$
v = \mu \;, \quad \tilde{w} = \nu \;, \quad u = \phi \;,
$$
where in the r.h.s. the standard notation for the CAT fields is used.
So one may expect that some
relationships exist
between our realization of $\w$ and the infinite-dimensional
symmetries of the CAT model found in \cite{{a3a},{a3b}}.

Let us remind
that the basic objects in the CAT model are the stress-tensor $W^2(z)$
and the spin 1 current $J^c(z)$. If the current $J^c(z)$ has a non-zero
central charge $c_J$ with respect to $W_2(z)$ and commutes with itself,
i.e. obeys the following OPE's
\begin{eqnarray}
W_2(z_1)J^c(z_2) & = & \frac{J^c(z_2)}{z_{12}^2} -
      \frac{\partial J^c(z_2)}{z_{12}} +  \frac{c_J}{z_{12}^3} \nn \\
J^c(z_1)J^c(z_2) & = &  0 \;,\label{d7}
\end{eqnarray}
then it is possible to construct the infinite number of currents
generating the higher spin symmetries of this model \cite{{a3a},{a3b}}. It
has been shown
in \cite{a3a} that in a special limit this symmetry is reduced to $\w$.

In our realization we have an analog of $J^c$, the current
$\hat{W}^{(-1)}_1 = \partial v$, but it has a vanishing central charge
with respect to the stress-tensor $2V^{(0)}$.
However, if we add to the set of our $\hat{{\cal G}}$ currents
one more set of higher spin currents
\be
A^{(s)} = (s+1) (\partial v)^s \; \partial^2 \tilde{w}\;,
\ee
which also generate symmetries of the
action (5.1), we become able to construct the CAT stress-tensor as a
linear combination
of the currents $V^{(0)}$ and $A^{(0)}$ at $\alpha = -\frac{1}{2}$
\be
W^2 (z)= 2 V^{(0)}(z)- 2 A^{(0)}(z)= 2\partial v \partial \tilde{w}
+(\partial u)^2 -\partial^2 u- 2\partial^2 \tilde{w} \;.
\ee
With respect to this stress-tensor, the current $\hat{W}^{(-1)}_1$ behaves as
a quasi-primary field with nonvanishing $c_{J}$ and so we recover the CAT
current structure on the spin 1 and spin 2 levels. Nevertheless it remains
not quite clear how our factor-algebra realization of $\w$ could be
incorporated into the CAT higher spin algebra. We have checked that
the realization of $\w$ following from the CAT algebra in the large $J^C$
limit \cite{a3a}
is different from ours: it does not involve the field $u (=\phi)$ and is
rather reduced to the two-field realization discussed in Sect. 3 and
Subsect. 4.1 (with some $\gamma \neq -\frac{1}{2}$). Perhaps, in order to
clarify these points one needs to consider a nonlinear realization of the
extension of $\hat{G}$ by the currents $A^{(s)}$. One of the most
interesting problems in this context is how to obtain the CAT field equations
starting from some nonlinear realization and applying the covariant reduction
procedure (i.e.
implementing a dynamical version of the inverse Higgs constraints
\cite{{ab},{a7},{ac}}).

\setcounter{equation}{0}
\section{Summary and discussion}

In this paper we discussed how the nonlinear (coset space) realization
approach of ref.\cite{a0} could be extended to produce multi-field
realizations of $\w$.
We considered two possibilities, one of which is to narrow the stability
subalgebra (respectively, to enlarge the coset space) still staying with
the original $\w$ and the other is to embed $\w$ into a larger symmetry
and to constuct a nonlinear realization of the latter. In both cases
we found non-trivial examples of multi-field realizations of $\w$. We
reproduced the two-field realization of $\w$ \cite{a2} as a coset space one
corresponding to the extension of $\w$ by one more set of
self-commuting higher spin
currents, i.e. to
the algebra $\hat{{\cal G}}$. Further we constructed a new
dilaton-like two-field realization of
$\w$ in the coset space involving all the spin 2 (Virasoro) generators.
In order to find a reasonable invariant action for this system it turned
out necessary to add one more field with a specific transformation law
under $\w$. We have shown that this new three-field realization corresponds
to embedding $\w$ as a factor-algebra into some nonlinear deformation
$\hat{{\cal G}}$ of the previously employed algebra $\tilde{{\cal G}}$.
This new algebra can as well be regarded as an extension of the universal
enveloping of some commuting centreless $U(1)$ Kac-Moody algebra, so that
this enveloping forms an ideal while the factor-algebra by this ideal
is $\w$. We proposed some generalizations of the three-field realization
to include more fields. One of them is the four-field realization with
the signature $(2+2)$ in the target space, such that the extended
algebra $\hat{{\cal G}}$ is reduced for it to the standard $\w$. This system
is expected to have an intimate relation to the $N=2$ string \cite{a9} and to a
$N=2$ supersymmetrization of the two-field realization of ref. \cite{a2}.
Finally, we discussed possible parallels between our three-field system and the
affine conformal Toda systems \cite{{a3},{a3a},{a3b}}.

There remain many unsolved problems in the coset space approach to $\w$.
First of all, it is as yet unclear how to reproduce in this framework the
most interesting multi-field realizations of $\w$ \cite{aa}
which basically correspond to the replacement of the singlet field
in the one-field realization by some matrix field taking values
in the fundamental representation of some classical algebra or its Cartan
subalgebra. In the light of the discussion in the present paper, it seems that
the only way to do this is to embed $\w$ into some extended algebra which would
contain as many spin 1 generators as the fields in the given realization, in
order to enable us to interpret these fields as coset parameters. A link with
the classical Lie algebras could arise as the property that the spin 1
generators,
being mutually commuting, transform according to the
fundamental representations
of these algebras, treated as some outer automorphism ones (as in
extended supersymmetries). It is very intriguing to reveal what such huge
higher spin algebras could be. If existing, they could have innumerable
implications in the theory of $W$ strings, $W$ gravities, etc.

Regarding the realizations presented in this paper, there are also some
points about them which need a further clarification. Some of these
issues were already discussed in the main body of the paper. Here we wish to
fix one more interesting problem, concerning the geometric interpretation
of the extended symmetries explored in this paper. The algebra $\w$ has a
nice interpretation as the algebra of the volume-preserving diffeomorphisms
of a cylinder \cite{{a0},{aa}}. It
is desirable to have analogous geometric images for the algebras
$\tilde{{\cal G}}$ and $\hat{{\cal G}}$. While the first algebra can be
considered as a contraction of a direct sum of two algebras $\w$ and so can
hopefuly be understood from the point of view of two independent
diffeomorphism groups, it remains a mystery what is the geometric meaning of
the algebra $\hat{{\cal G}}$. Perhaps it could be somehow related (e.g. through
a contraction) to the recently discovered algebra $\hat{W}^{\infty}$ \cite{a11}
which is claimed to contain all possible $W$ type algebras, either via a
contraction or a truncation.

Finally, we wish to point out that the obvious interesting problem ahead is
to gauge
the above realizations of $\w$. This might clarify their relation to
$w_{\infty}$ strings and $w_{\infty}$ gravity.

\def\thesection { }
\section{Acknowledgements}

Two of us (E.I.\& S.K.) are very grateful to K.Stelle for bringing
to their attention the preprint \cite{a0} before publication and
for illuminating discussions at the early stage of the present study.
They also wish to thank
the INFN for financial support and the Laboratori Nazionali di Frascati for
hospitality extended to them during
the course of this work. E.I. thanks ENSLAPP in Lyon for hospitality.

\end{document}